\documentclass[twocolumn,prl,amsmath,amssymb,showpacs,nofootinbib]{revtex4}

\def\lsim{\, \rlap{$<$}{\lower 1.1ex\hbox{$\sim$}}\,}

\begin{document}

\title{Scaling Solution For Small Cosmic String Loops}

\author{Jorge V. Rocha}
\email[]{jrocha@physics.ucsb.edu}
\affiliation{Department of Physics, University of California, Santa Barbara, California 93106, USA}

\begin{abstract}
The equation governing the time evolution of the number density of loops in a cosmic string network is a detailed balance determined by energy conservation.  We solve this equation with the inclusion of the gravitational radiation effect which causes the loops to shrink (and eventually decay) as time elapses.  The solution approaches a scaling regime in which the total energy density in loops remains finite, converging both in the infrared and in the ultraviolet.
\end{abstract}


\maketitle

\section{\label{Intro}Introduction}

More than 30 years ago it was realized that a network of one-dimensional topological defects could form during a cosmological phase transition~\cite{Kibble:1976sj}.  Such {\em cosmic strings} consequently evolve in a fashion determined by the expansion of the universe, intercommutation events and gravitational radiation.  All these mechanisms act in such a way so as to yield a scaling regime, in which the network remains self-similar under rescaling, at least in its large scale properties.  This has been convincingly established by now, both through numerical~\cite{Albrecht:1989mk,Allen:1990tv,Bennett:1989yp,Ringeval:2005kr,Martins:2005es,Olum:2006ix} and analytic~\cite{Kibble:1990ym,Austin:1993rg} studies, and it is widely believed that the inter-string distance and the correlation length along a long string both scale.  However, the existence or not of a scaling regime for the smaller scales (for example, the typical size of the small loops) still remains an open question~\cite{Polchinski:2007qc}.  It is this problem we wish to address in this short note.  Understanding the properties of small cosmic loops would be a valuable achievement: the detectability of gravitational waves emitted by cosmic strings~\cite{Damour:2004kw,Siemens:2006yp,Dubath:2007wu} depends crucially on their knowledge, while the size of the smallest loops can also have some effect on microlensing~\cite{Chernoff:2007pd}.  To this end we will consider the number density of loops -- it is the small scale structure that is responsible for their production and so the number density of small loops provides a measure of the former.  Presently, we demonstrate that this quantity scales.

There are many processes that can act to change the number density of loops with a given length $l$.  Expansion of the universe causes the strings to stretch, but only on scales larger than the horizon size~\cite{Vilenkin:1981kz}.  Therefore loops essentially do not grow in time.\footnote{Loops bigger than the horizon distance are regarded as long strings.}  Self-intersections of cosmic strings can produce loops from long strings and also fragment loops into smaller ones.  This mechanism was studied in~\cite{Polchinski:2006ee}, taking as an input the simplified analytic model developed therein, which we believe describes fairly accurately the evolution of the small scale structure on cosmic string networks.  Of course, string intercommutation can lead to the absorption of loops back into the long string population but this process is strongly suppressed for small loops and so can be neglected.  Finally, the coupling of matter in the form of cosmic strings to gravity means that the network radiates away part of its energy.  This has been amply discussed in the literature~\cite{Turok:1984cn,Vachaspati:1984gt,Garfinkle:1987yw,Siemens:2001dx} and as a consequence loops shrink as time progresses and eventually disappear from the network.

The equation governing the evolution of the number density of loops in a cosmic string network is a detailed balance determined by conservation of energy.  In what follows we solve this equation, first ignoring gravitational radiation and then taking into account this effect.  It is shown that the solution approaches a scaling regime and if we include GW emission the total energy density in loops remains finite, in particular converging in the ultra-violet (UV), even when the loop production function diverges.  Above the gravitational radiation scale the loop energy density decreases with the size of the loops as a power-law which is in good agreement with the numerical simulations of~\cite{Ringeval:2005kr,Olum:2006ix} and implies the convergence of the total energy density in loops in the infra-red (IR).  However, below the gravitational radiation scale this quantity falls off with a lower power of the loop length in such a way that the energy density remains finite in the UV limit also.  Finally, we comment on the curious fact that this sub-gravitational radiation regime seems to be apparent in the results of~\cite{Ringeval:2005kr}, even though those simulations did not include gravitational radiation.

\section{\label{Prelim}Preliminaries}

Several scales can be defined for a cosmic string network.  Among them, the {\em characteristic length} of the network plays an important part.  It is defined as the length scale $L$ such that a typical volume $L^3$ of the network contains a length $L$ of long strings.  Another important concept is that of a {\em scaling regime}.  We say that a quantity (with units of length) is scaling if it remains constant in units of the cosmological time $t$ as the network evolves.  Thus, in a scaling regime we have $L=\gamma t$, for some constant $\gamma$, and the energy density in long strings becomes
\begin{equation}
\rho_\infty = \frac{\mu L}{L^3} = \frac{\mu}{\gamma^2 t^2} \ ,
\end{equation}
where $\mu$ is the string tension.  According to Ref.~\cite{Ringeval:2005kr} the constant $\gamma^{-2}$ takes the value $\sim 3$ in a matter-dominated universe and $\sim 9.5$ in a radiation-dominated universe, whereas Ref.~\cite{Martins:2005es} provides the values $\gamma^{-2} \sim 3$ and $\gamma^{-2} \sim 11.5$, respectively. 

Now, denote by $\frac{dn}{dl}(l,t)$ the number density of loops with size comprised between $l$ and $l+dl$.  This quantity has units of $({\rm length})^{-4}$.  Thus, if it ever reaches a scaling regime during the cosmological evolution, it must eventually approach the following form:
\begin{equation}
\frac{dn}{dl} = t^{-4} f(l/t) \ .
\end{equation}
Borrowing notation from~\cite{Ringeval:2005kr}, define the length of loops in units of the horizon size, $\alpha\equiv l/d_{\rm h}$.  If the scale factor takes the form $a(t)\propto t^\nu$, the horizon size can be expressed as $d_{\rm h}=t/(1-\nu)$ and so the signature of a scaling regime in the loop number density is a solution of the form
\begin{equation}
\frac{dn}{d\alpha} = \frac{{\cal S}(\alpha)}{\alpha \, d_{\rm h}^3} \ .
\label{scaling}
\end{equation}
${\cal S}(\alpha)$ is known as the {\em scaling function}.

In what follows we shall need expressions for the rate at which the loops shrink due to emission of gravitational radiation and for the rate of loop formation.  The power radiated away by cosmic strings is typically given by $P = \Gamma G\mu^2$, where $G$ is Newton's gravitational constant and $\Gamma$ is a numerical constant of order 50~\cite{Vachaspati:1984gt,Burden:1985md,Garfinkle:1987yw,Quashnock:1990wv,Allen:1994bs}.  Since a loop of length $l$ has energy $\mu l$, this implies that loops shrink at a rate
\begin{equation}
\frac{dl}{dt} = - \Gamma G\mu \ .
\label{gravrad}
\end{equation}

A first attempt to analytically estimate the loop production from long strings~\cite{Polchinski:2006ee} resulted in a divergent total energy density in small loops.  There it was found that the average number of loops produced per unit time, per unit distance along the string and per unit loop length is given by
\begin{equation}
\frac{d\left<{\cal N}\right>}{dt \, d\sigma \, dl} = \frac{c}{l^3} \left(\frac{l}{t}\right)^{2\chi} \ ,
\label{loopprod}
\end{equation}
where $c$ is the overall normalization of the loop production function.  For a matter-dominated era $\chi=0.25$ and $c=0.042$, while during the radiation epoch $\chi=0.10$ and $c=0.121$.  These values of the normalization $c$ are most likely to be over-estimates since fragmentation was not taken into account.  Indeed, these values of $c$ exceed the numerical results of~\cite{Ringeval:2005kr} by factors of approximately 15 and 40, while comparison with~\cite{Olum:2006ix} yields factors of 3.5 and 85, respectively.  The latter reference considered the loop production function and so their results relate more directly to the normalization $c$, whereas the former reference studies the loop number density.  Note that even though these simulations might seem in conflict, the authors of~\cite{Olum:2006ix} emphasize that the range of lengths corresponding to the small loops is not yet scaling.  Furthermore, the determination of the normalization from the scaling function is sensitive to the choice of the range where the fitting is performed in~\cite{Ringeval:2005kr}.  So there is a possibility that there exists no real discrepancy between both simulations.  

For a consistent description of the network, equation~(\ref{loopprod}) must be corrected at small scales.  The correction is provided by the smoothing due to emission of gravitational waves (GW) from cosmic strings and has been computed in~\cite{Siemens:2002dj} considering a discrete spectrum for the fluctuations.  More recently an improved result performed in the continuum limit was obtained in~\cite{Polchinski:2007rg}, where the structure along the strings was found to be smoothened on scales below 
\begin{equation}
l_{GW} \approx 20 (G\mu)^{1+2\chi} t \ .
\label{smoothing}
\end{equation}
Hence, gravitational radiation introduces a natural cutoff for the divergent loop production function, even though it is believed to be unnecessary to achieve scaling~\cite{Ringeval:2005kr,Martins:2005es}.  However, we will disregard this effect in what follows and see how far we can get.

\section{Evolution of the loop number density neglecting gravitational radiation}

As discussed in~\cite{Polchinski:2006ee}, if gravitational radiation is neglected the number of loops within a comoving volume changes only due to loop production:
\begin{equation}
\frac{d}{dt} \left(a^3 \frac{dn}{dl}\right) = \frac{c\, a^3}{\gamma^2 t^2 l^3} \left(\frac{l}{t}\right)^{2\chi} \ ,
\end{equation}
Defining $F(l,t)\equiv\frac{dn}{dl}(l,t)$ and inserting the power-law expression for the scale factor we obtain
\begin{equation}
t \dot F + 3\nu F = \frac{c}{\gamma^2 l^4} \left(\frac{l}{t}\right)^{1+2\chi} \ ,
\label{evol}
\end{equation}
which has the general solution
\begin{equation}
F(l,t) = \frac{c}{\gamma^2(3\nu - 1 - 2\chi) t^4} \left(\frac{l}{t}\right)^{2\chi-3} + \frac{C_0(l)}{t^{3\nu}} \ .
\end{equation}
The function $C_0$ depends only on the variable $l$.  Noting that the inequality $1+2\chi < 3\nu$ is satisfied both in the radiation- and matter-dominated eras, we conclude that at late times ($t\rightarrow\infty$) the loop number density approaches
\begin{equation}
\frac{dn}{dl}(l,t) \longrightarrow \frac{c}{\gamma^2(3\nu - 1 - 2\chi) t^4} \left(\frac{l}{t}\right)^{2\chi-3} \ .
\end{equation}
Thus, expressing everything in terms of $\alpha$ and $d_{\rm h}$ we indeed find that the loop number density approaches a scaling regime, i.e. it takes the form~(\ref{scaling}) with 
\begin{equation}
{\cal S}(\alpha) = \frac{c\, (1-\nu)^{-1-2\chi}}{\gamma^2 (3\nu-1-2\chi)} \alpha^{2\chi-2} \ .
\label{sol1}
\end{equation}
This confirms that the loop number density does approach a scaling regime without taking into account gravitational radiation.  The exponent $2\chi-2 \equiv -p$ takes the values $-1.5$ in the matter era and $-1.8$ in the radiation era.  This is in good agreement with the numerical results of~\cite{Ringeval:2005kr} who quote $p_{mat} = 1.41^{+0.08}_{-0.07}$ and $p_{rad} = 1.60^{+0.21}_{-0.15}$.  However, these simple power-laws become good fits to the data only above a physical length $\ell_{\rm c}$, which is identified with the initial correlation length of the network.  Furthermore, if the solution~(\ref{sol1}) were valid over the full range of loop lengths the total energy density would diverge in the UV since
\begin{equation}
\int l \frac{dn}{dl} dl = d_{\rm h}^{-2} \int {\cal S}(\alpha) d\alpha \ .
\label{total}
\end{equation}
We will now show that including the process of shrinkage of the loops due to gravitational radiation changes the power-law below the gravitational radiation scale, thus yielding a convergent total energy density in loops.

\section{Evolution of the loop number density including gravitational radiation}

Inclusion of gravitational radiation into the evolution equation for the loop number density introduces an extra term on the left-hand side of equation~(\ref{evol}) because loops shrink at a rate given by~(\ref{gravrad}).  Therefore, we now have
\begin{equation}
t \dot F + 3\nu F - \Gamma G \mu F' = \frac{c}{\gamma^2 l^4} \left(\frac{l}{t}\right)^{1+2\chi} \ .
\end{equation}
Defining for convenience $b \equiv \Gamma G \mu$, the solution of the above differential equation may be written as
\begin{eqnarray}
F(l,t) & = & \frac{c\, b}{\gamma^2} \frac{(l+bt)^{2\chi-4}}{t^{2\chi}} \left(\frac{bt}{l+bt}\right)^{2\chi-3\nu} \nonumber \\
& \times & B\left( \frac{bt}{l+bt}; 3\nu-1-2\chi, 2\chi-2 \right) \nonumber \\
& + & \frac{C(l+bt)}{t^{3\nu}} \ ,
\label{sol2}
\end{eqnarray}
where $B$ represents the Euler incomplete beta function and $C$ can be any function of the combination $l+bt$.

By employing the Taylor expansion
\begin{equation}
B\left( \epsilon; 3\nu-1-2\chi, 2\chi-2 \right) = \frac{\epsilon^{3\nu-1-2\chi}}{3\nu-1-2\chi} + O(\epsilon)\ ,
\label{taylor}
\end{equation}
we find that the first term in~(\ref{sol2}) behaves, for large $l$ and fixed $t$, as $\sim l^{2\chi-3}$, whereas the second term goes like $\sim C(l)$.  Requiring that the energy density in loops converges in the limit $l \rightarrow \infty$ imposes that the general function $C(x)$ decays faster than $x^{-2}$.  Therefore, using the expansion
\begin{eqnarray}
\lefteqn{B \left( 1-\epsilon; 3\nu-1-2\chi, 2\chi-2 \right) =} \nonumber \\
& - & \, \frac{\pi\csc(2\pi\chi)\Gamma(3\nu-2\chi)\epsilon^{2\chi-2}}{(3\nu-1-2\chi)\Gamma(2\chi-1)\Gamma(3-2\chi)\Gamma(3\nu-1-2\chi)}  \nonumber \\ 
& + & \, O\left(\frac{1}{\epsilon}\right) \ ,
\label{taylor2}
\end{eqnarray}
the second term in~(\ref{sol2}) is dominated by the first term as $t \rightarrow \infty$:
\begin{eqnarray}
\frac{dn}{dl}(l,t) & \longrightarrow & \frac{c\, b}{\gamma^2} \frac{(l+bt)^{2\chi-4}}{t^{2\chi}} \left(\frac{bt}{l+bt}\right)^{2\chi-3\nu} \nonumber \\
& \times & B\left( \frac{bt}{l+bt}; 3\nu-1-2\chi, 2\chi-2 \right) \ .
\end{eqnarray}
Once again converting to the variables $\alpha$ and $d_{\rm h}$, we obtain a solution of the form~(\ref{scaling}) with
\begin{eqnarray}
\lefteqn{{\cal S}(\alpha) = \frac{c\, b^{2\chi-3} \alpha}{\gamma^2 (1-\nu)^4} \left(\frac{\alpha+(1-\nu)b}{(1-\nu)b}\right)^{3\nu-4}} \nonumber \\
& \times & B\left( \frac{(1-\nu)b}{\alpha+(1-\nu)b}; 3\nu-1-2\chi, 2\chi-2 \right) \ .
\end{eqnarray}

Now we can use the series expansions~(\ref{taylor}) and~(\ref{taylor2}) to recover the limits for small and large loops.  The separation between these two regimes is set by the gravitational radiation scale $\Gamma G \mu$, and we find for $\alpha \gg \Gamma G \mu$ the same result~(\ref{sol1}), whereas for $\alpha \ll \Gamma G \mu$
\begin{eqnarray}
{\cal S}(\alpha) & \simeq & \frac{\pi(1-2\chi)\csc(2\pi\chi)\Gamma(3\nu-2\chi)}{(3\nu-1-2\chi)\Gamma(2\chi)\Gamma(3-2\chi)\Gamma(3\nu-1-2\chi)} \nonumber \\
& \times & \frac{c}{\gamma^2 \Gamma G \mu (1-\nu)^{2+2\chi}} \alpha^{2\chi-1} \ .
\end{eqnarray}
Since $0 < 2\chi < 1$ holds in both cosmological eras, the integral~(\ref{total}) is manifestly convergent and so the total energy density in loops is finite, as desired.

\section{Discussion}

We may now compare our results with those obtained numerically in~\cite{Ringeval:2005kr}.  We have already noted that the exponent $2\chi-2$ agrees well with the simulations for the larger loops.  However, we have shown that taking into account gravitational radiation has the effect of bending the curves for $\alpha \lsim \Gamma G \mu$ so that the exponent then becomes $2\chi-1$.  Indeed, Figure~3 of that reference does appear to show a certain range of the parameter $\alpha$ in which the scaling function behaves as such a power-law.  At first sight this might seem intriguing since those simulations did not include the gravitational radiation process directly.  However, small loops behave like matter and so the expansion of the universe effectively shrinks the loops.  Because the simulations only keep loops with sizes greater than a fixed fraction of the horizon, they are eventually removed from the game, hence emulating gravitational radiation\footnote{I thank F. Dubath for pointing out this fact to me.}.  The minimum counting size was $\alpha_{min} = 10^{-5}$ so we should expect the sub-gravitational radiation regime to set in for comparable scales or smaller.  A more accurate estimate, equating the lifetime of the loops determined by gravitational decay to the lifetime set by the minimum counting size, yields $\alpha_{bend} = \frac{2-\nu}{1-\nu}\alpha_{min}$ for the scale at which the bending would occur.  Nonetheless, it is curious that this sub-gravitational radiation regime shows up also at early times in the simulations, when scaling is yet to be achieved.

We also note that the inclusion of gravitational radiation in our equations leads to a finite energy density of loops, even though the loop production function used as an input diverges at small scales.  This means that the rate at which loops are removed from the network is sufficiently high to balance the diverging loop formation.  Of course, this divergence must be eliminated in order to satisfy string length conservation.  The gravitational smoothing lengthscale~(\ref{smoothing}) naturally introduces a UV cutoff on the loop production function and so this should be the typical size of cosmic loops, as advocated in~\cite{Polchinski:2007rg}.  It is comforting that this cutoff sets in much before the scale of validity of the effective theory, namely the thickness of the strings~\cite{Vincent:1996rb}, is reached, as long as the value of the string tension is not too small.  It remains to be seen if the the correct normalization of the loop production function can be reproduced analytically by methods similar to those used in~\cite{Polchinski:2006ee}.  This is left for future work.

\section{Acknowledgments}

I thank Joseph Polchinski and Florian Dubath for useful discussions.  I am also grateful to Joseph Polchinski for reading a draft of this paper and to Mark Wyman for encouragement.  I acknowledge financial support from Funda\c{c}\~{a}o para a Ci\^{e}ncia e a Tecnologia, Portugal, through Grant No. SFRH/BD/12241/2003.

\bibliography{scaling_loops_bib}

\end{document}